\documentclass[a4paper, 12pt]{article}

\usepackage[left=1.5cm, right=1.5cm, top=1.5cm]{geometry}
\usepackage{amsmath}
\usepackage{amssymb}
\usepackage{graphicx}% Include figure files
\usepackage{subfig}
\usepackage{dcolumn}% Align table columns on decimal point
\usepackage{bm}% bold math
\usepackage[utf8]{inputenc}
\usepackage[T1]{fontenc}
\usepackage{mathptmx}
\usepackage{etoolbox}
\usepackage[square,numbers]{natbib}

\usepackage{enumitem}

\title{Neutral pion momentum in hypertriton mesonic decay through a root-finding method}

\author{
Emile Meoto\\[0.6cm]
Department of Physics, University of Buea\\
P.~O.~Box 63, Buea, South West Region, Cameroon\\
Email: meoto.emile@ubuea.cm or emeotoson@gmail.com
}

\date{\today}

\begin{document}

\maketitle

\begin{abstract}
A root-finding method is used to study two-body mesonic decay in the hypertriton. We validate this Newton--Raphson root-finding approach by applying it to the negative-pion decay channel ($^{3}_{\Lambda}\mathrm{H} \rightarrow {}^{3}\mathrm{He} + \pi^{-}$), for which the pion momentum and lambda binding energy were recently reported by MAMI A1 Collaboration as $p_{\pi^-} = 113.789 \pm 0.020_{\text{stat.}} \pm 0.112_{\text{syst.}} \text{ MeV}/c$ and $B_{\Lambda} = 0.523 \pm 0.013_{\text{stat.}} \pm 0.075_{\text{syst.}}$ MeV, respectively. Using their reported $\Lambda$ binding energy, the root-finding method and an exact kinematic formula both yield $p_{\pi^{-}} = 113.790$ MeV/$c$, agreeing with each other. We then apply both the Newton--Raphson method and the exact formula to the neutral-pion decay channel ($^{3}_{\Lambda}\mathrm{H} \rightarrow {}^{3}\mathrm{H} + \pi^{0}$), for which the neutral pion momentum cannot be directly measured due to difficulties in experimental setup. Both methods agree, yielding a predicted neutral-pion momentum of $p_{\pi^{0}} = 118.129$ MeV/$c$. This validates the root-finding algorithm as a robust equivalent for predicting pion momenta that may be experimentally inaccessible in some cases. Furthermore, it establishes the method as a reliable tool for extension to three-body mesonic decays, for which the pion momentum is a continuum and the exact kinematic formula can no longer be applied. In addition, the pion momentum computed allows for its 4-momentum to be completely determined, a useful input for investigating its two-photon decay ($\pi^0 \to \gamma \gamma$) in the rest frame of the hypertriton.
\end{abstract}

\section{Introduction}

The accurate determination of the structural properties of the hypertriton is currently receiving significant attention in a number of experiments. This high interest in the hypertriton stems from its fundamental role in hypernuclear physics as the lightest hypernucleus, and in particular, its role in constraining the lambda-nucleon interaction in the absence of reliable scattering data. Some of the observables measured in experiments include its lifetime, binding energy, production cross sections and yields, spin structure and parity. Some of these observables are currently being refined in ongoing experiments while others are the subject of planned experiments. One of the methods that is used to probe the hypertriton is decay pion spectroscopy. In this method, the momentum of a pion from two-body mesonic decay of the hypertriton is measured and used to compute binding energy. 

Recently, the Mainz Microtron (MAMI) A1 collaboration published a new lambda binding energy of the hypertriton that was determined through decay pion spectroscopy. Their reported value is $B_\Lambda(^3_\Lambda\mathrm{H}) = 0.523 \pm 0.013_{\rm stat.} \pm 0.075_{\rm syst.}
\ \mathrm{MeV}$. The decay channel used in this experiment is the two-body negative pion decay

\begin{equation}
{}^{3}_{\Lambda}\mathrm{H}
\rightarrow
{}^{3}\mathrm{He} + \pi^- .
\end{equation}

In order to determine the hypertriton binding energy, the pion momentum ($p_{\pi^-}$) from this reaction was first measured. From this measured momentum, the hypertriton invariant-mass is reconstructed, and then the binding energy is calculated.

Another important two-body decay channel is the neutral pion decay.

\begin{equation}
{}^{3}_{\Lambda}\mathrm{H}
\rightarrow
{}^{3}\mathrm{H}
+\pi^{0}.
\end{equation}

The case of neutral pion decay is completely different: there is currently no experimental measurement of the neutral pion momentum from this decay of the hypertriton. Neutral pion decays are significantly more difficult to observe directly than charged pion. This difficulty arises from the fact that A neutral pion has a shorter lifetime, and decays to two photons through the electromagnetic interaction $\pi^0 \to \gamma  \gamma$. Secondly, they carry no charge. Charged pions that cause ionisation, leaving clear tracks in detectors and their momentum can be precisely measured from the curvature of these tracks in a magnetic field. Since neutral pions produce no ionization tracks, they are invisible to standard charged-particle trackers. Detection relies entirely on their decay products i.e. two photons. These photons may be detected in electromagnetic calorimeters \cite{kor2000}. 

Hypertriton mesonic-decay literature is not huge. It may be organised into two clear cases: theoretical calculations on lifetime and branching-ratio from various two and three-body mesonic channels, and experiments. Long before the few-body calculations in Refs. \cite{gon1992, glo1998, kam1998}, early foundations on theoretical work were laid down in Refs. \cite{dal1958, ray1966} that were dedicated to studying the hypertriton through phenomenology. Recent theoretical calculations are reported in Refs. \cite{gal2019, per2020, san2021, hil2023}. As discussed in Ref. \cite{hil2023}, accurate pion momenta are a crucial input to calculations that quantify the role of final-state interactions on the mesonic decay. Experimental results that measure observables based on mesonic decay channels include Refs \cite{ali2019,ada2020, kin2026}.

In this paper, we make use of the binding energy from the MAMI A1 collaboration \cite{kin2026} and determine the pion momentum from neutral pion decay through the Newton-Raphson algorithm. This methodology was developed in Ref.~\cite{meo2026}. It is important to emphasise that the pion momentum can be determined exactly for two-body decay channels because it is monochromatic \cite{kam1998}. Our methodology of finding this pion momentum through a root-finding algorithm becomes indispensable for three-body mesonic decays where the momentum is a continuum. The present two-body study
therefore serves as an essential testing ground for the method before it is extended to those more complex three-body channels, as demosntrated in Ref. \cite{meo2026}.

\section{Mesonic decay in hypertriton}

In free space, the dominant mesonic decay modes of the \(\Lambda\) hyperon are negative-pion and neutral-pion decays.

\begin{align}
\Lambda &\to p + \pi^- \\
\Lambda &\to n + \pi^0
\end{align}

where $m_pc^2$, $m_{\pi^-}c^2$, $m_nc^2$, and $m_{\pi^0}c^2 $ are the masses of the proton, negative pion, neutron and neutral pion, respectively \cite{pdg2024, cod2022}. The Q-values for these free decays are $Q_{\pi^-}=37.841 MeV $ and $Q_{\pi^0}=41.141 MeV $. Inside a hypernucleus $_{\Lambda}^{3} H$, mesonic decays take place and the effective Q-values for these mesonic decays in nuclear environment are modified from the values in free space. As derived in Ref. \cite{meo2026}, for negative pion and neutral pion decay, these effective Q-values are given, respectively, by

\begin{subequations}
\begin{align}\label{eq:proton_eff} 
Q_{\pi^-}^{\mathrm{eff}}=Q_{\pi^-} - B_\Lambda + S_p, \\
\label{eq:neutron_eff}
Q_{\pi^{0}}^{\mathrm{eff}} = Q_{\pi^{0}} - B_{\Lambda} + S_{n},
\end{align} 
\end{subequations}

$S_p=5.4934$ MeV is the proton separation energy of ${}^{3}\mathrm{He}$ while $S_n=6.257$ MeV is the neutron separation energy of ${}^{3}\mathrm{H}$. The structure of the relations in Eqs. \ref{eq:proton_eff} and \ref{eq:neutron_eff} reveals a very important aspect of pion spectroscopy: the lambda binding energy $B_\Lambda$ presents a direct link between negative and neutral pion decays. Through this link, a high-precision measurement in any mesonic decay channel, opens up the possibility to study all other decay channels (both two-body and three-body decay channels). This is very important because some channels are extremely difficult to manage in experimental setups. 

Since $S_p>0$, ${}^{3}\mathrm{He}$ is stable against proton decay. Similarly, $S_n>0$ implies that ${}^{3}\mathrm{H}$ is stable against neutron decay.  At this stage, we make use of the $B_\Lambda$ from the MAMI A1 experiment to determine the effective Q values: 

\begin{align}
Q_{\pi^-}^{\text{eff}} &= 37.841 - 0.523  + 5.494 = 42.812~\text{MeV}.\\
Q_{\pi^0}^{\text{eff}} &= 41.141 - 0.523  + 6.257 = 46.875~\text{MeV}.
\end{align}

This is the key quantity that is needed to set up the polynomial equations for the pion momentum.

%%%%%%%%%%%%%%%%%%%%%%%%%%%%%%%%%%%%%%%%%%%%%%%%%%%%%%%%%%%%%%%%%%%%%%%%%%%%%%%%%%%%%%%%%%
\section{Results and discussion}

As stated at the outset, the goal of this paper is to compute neutral pion momentum in neutral pion mesonic decay. The negative pion momentum from negative pion decay was recently measured by the MAMI A1 collaboration as $p_{\pi^-}= 113.789 \pm 0.020_{\text{stat.}} \pm 0.112_{\text{syst.}} \text{ MeV}/c$. This is the value that was used to compute the value of $B_{\Lambda}=0.523$. As a demonstration of the accuracy of our method, we apply the method to recover the value of $p_{\pi^-}=113.789$ MeV/c from experiment, before using the method to predict a value for $p_{\pi^0}$. In addition to recovering the experimental value of $p_{\pi^-}$, we also show how the available energy is shared between the decay products i.e. we compute the recoil energy of ${}^{3}\mathrm{He}$. 

\subsection{Negative pion decay of hypertriton}

For the two-body negative-pion decay of $^3_\Lambda$H, 4-momentum conservation in the rest frame of the hypertriton gives rise to the following relations:

\begin{equation}
Q_{\pi^-}^{\text{eff}} = T_{^3\text{He}} + T_{\pi^-},
\end{equation}
\begin{equation}
\vec{0} = \vec{p}_{^3\text{He}} + \vec{p}_{\pi^-},
\end{equation}
where
\begin{equation}
T_{^3\text{He}} = \sqrt{(M_{^3\text{He}}c^2)^2 + (p_{\pi^-}c)^2} - M_{^3\text{He}}c^2,
\end{equation}
\begin{equation}
T_{\pi^-} = \sqrt{(m_{\pi^-}c^2)^2 + (p_{\pi^-}c)^2} - m_{\pi^-}c^2.
\end{equation}

Substituting $Q_{\pi^-}^{\text{eff}} = 42.812$~MeV, $M(^3\text{He})c^2 = 2808.392$~MeV, and $m_{\pi^-}c^2 = 139.570\,39$~MeV into the energy-conservation equation and letting $p_{\pi^-} \equiv |\vec{p}_{\pi^-}|$,

\begin{equation}
42.812 = \sqrt{2808.392^2 + p_{\pi^-}^2c^2} - 2808.392 + \sqrt{139.570\,39^2 + p_{\pi^-}^2c^2} - 139.570\,39,
\end{equation}

We then search for the root of this equation through the Newton-Raphson method. This method locates the root at $p_{\pi^-} = 113.790$~MeV/c. The experimental value measured at MAMI A1 experiment is $p_{\pi^-} = 113.789$~MeV/c. Our computed value has a relative deviation of only 0.0009\% from the central value of the experiment. This confirms the accuracy of our root-finding method as applied to the mesonic decay problem. The convergence behaviour of the pion momentum is shown in Table \ref{tab:newton_convergence}.
.
\begin{table}[h]
\centering
\caption{Newton-Raphson convergence for the negative-pion momentum $p_{\pi^-}$.}
\label{tab:newton_convergence}
\begin{tabular}{cc}
\hline\hline
Iteration $n$ & $p_{\pi^-}$ (MeV/c) \\
\hline
0 & 144.86178429 \\
1 & 115.69058034 \\
2 & 113.79991751 \\
3 & 113.79022447 \\
4 & 113.79022421 \\
5 & 113.79022421 \\
\hline\hline
\end{tabular}
\end{table}

The collinear recoil momentum of $^3$He is $p_{^3\text{He}} = 113.790 $~MeV/c, in a direction opposite to that of the negative pion. The kinetic energy of each body is given by

\begin{align}
T_{\pi^-} &= \sqrt{(m_{\pi^-}c^2)^2 + (p_{\pi^-}c)^2} - m_{\pi^-}c^2 = 40.508~\text{MeV} \\
T_{^3\text{He}} &= \sqrt{(M_{^3\text{He}}c^2)^2 + (p_{\pi^-}c)^2} - M_{^3\text{He}}c^2 = 2.304  ~\text{MeV}
\end{align}

These results reveal that from the total available energy ($Q_{\pi^-}^{\text{eff}} = 42.812$~MeV), the kinetic energy given to the negative pion is 94.62\% of $Q_{\pi^-}^{\text{eff}}$. The recoil energy of $^3\text{He}$ is 5.38\% of $Q_{\pi^-}^{\text{eff}}$. This is the recoil energy of $^3\text{He}$ that is implied in the MAMI A1 measurement, even though it was not stated. 

It is instructive to see what impact the measured binding energies from other experiments have on the
pion momentum. Using the STAR measurement of $B_\Lambda = 0.406 \pm
0.120_{\rm stat.} \pm 0.110_{\rm syst.}$~MeV~\cite{ada2020} and the older emulsion measurement $B_\Lambda = 0.13 \pm 0.05$~MeV \cite{jur19731},  we repeat these calculations. Additionally, we cross-check each case (MAMI A1, STAR and emulsion) with the exact relativistic
kinematic formula of Ref.~\cite{kam1998}, which is also based on 4-momentum conservation. This exact formula is given by

\begin{align}
  k_\pi = \frac{\sqrt{\left(M_{\Lambda^3H}^2+M_{^3He}^2-m_\pi^2\right)^2-4M_{\Lambda^3H}^2M_{^3He}^2}}{2M_{\Lambda^3H}} .
\end{align}

\begin{table}[htbp]
\centering
\caption{Comparison of kinematic quantities for the negative pion decay
$^{3}_{\Lambda}\mathrm{H} \to {}^{3}\mathrm{He} + \pi^{-}$ using different $B_{\Lambda}$ measurements. The
row $p_{\pi^-}^{\rm formula}$ gives the pion momentum from the exact relativistic kinematic formula of
Ref.~\cite{kam1998}, for comparison with the Newton--Raphson result $p_{\pi^-}$.}
\begin{tabular}{lccc}
\hline\hline
Quantity & MAMI A1 ($B_{\Lambda}=0.523$ MeV) & STAR ($B_{\Lambda}=0.406$ MeV) & Emulsion ($B_{\Lambda}=0.13$ MeV) \\
\hline
$M_{{}^3_\Lambda\mathrm{H}}$ (MeV)          & 2990.773 & 2990.890 & 2991.166 \\
$Q_{\pi^-}^{\rm eff}$ (MeV)          & 42.812 & 42.929 & 43.205 \\
$p_{\pi^-}$ (MeV/$c$)               & 113.790 & 113.964 & 114.374 \\
$p_{\pi^-}^{\rm formula}$ (MeV/$c$) & 113.790 & 113.964 & 114.396 \\
$p_{^{3}\mathrm{He}}$ (MeV/$c$)     & 113.790 & 113.964 & 114.374 \\
$T_{\pi^-}$ (MeV)                   & 40.508 & 40.618 & 40.877 \\
$T_{^{3}\mathrm{He}}$ (MeV)         & 2.304  & 2.311  & 2.328  \\
\hline\hline
\end{tabular}
\label{tab:neg_pion_comparison_three}
\end{table}

The results are compared in Table~\ref{tab:neg_pion_comparison_three}.
It is immediately observed that the effective $Q$-value ($Q_{\mathrm{eff}}^{\pi^{-}}$) increases as $B_{\Lambda}$ decreases.
This follows directly from the relation
$Q_{\mathrm{eff}} = Q_{\pi^{-}} - B_{\Lambda} + S_p$: A lower lambda binding energy leaves more energy available to be shared by
the decay products. Therefore, the lower the lambda binding energy, the higher the pion momentum. The variation remains modest ($\sim 0.4$~MeV between MAMI A1 and the emulsion value). Both the pion momentum $p_{\pi^{-}}$ and the equal-and-opposite recoil momentum of ${}^{3}\mathrm{He}$
increase accordingly, rising from 113.790 MeV/$c$ (MAMI A1) to
114.374 MeV/$c$ (emulsion). The relative change is small ($\sim 0.5\%$),
indicating weak sensitivity of the pion momentum to the current range of $B_{\Lambda}$ uncertainties. Even
though older measurements had significantly lower $B_\Lambda$, the resulting changes in pion momentum and
recoil kinematics are minor. This highlights the robustness of decay-pion spectroscopy for constraining
hypernuclear properties, while also showing why high-precision measurements are invaluable for resolving
the hypertriton puzzle.

As Table~\ref{tab:neg_pion_comparison_three} shows, the Newton--Raphson momentum $p_{\pi^-}$ and the exact-formula momentum $p_{\pi^-}^{\rm formula}$ have excellent agreement across all three $B_\Lambda$
values. This confirms that the root-finding method reproduces the exact relativistic two-body kinematics. 

This agreement also lets us resolve a discrepancy with Ref.~\cite{kam1998}: that paper reports $\pi^{-}=117.4$
MeV/$c$ using a hypertriton mass derived from the emulsion measurement $B_\Lambda=0.13$ MeV. Yet, Table~\ref{tab:neg_pion_comparison_three} the pion momentum for this binding energy should be $p_{\pi^-} = 114.374$ MeV/$c$. It remains
unclear why the value reported in that paper is substantially higher, given the same input binding energy. 

\subsection{Consistency check of an alternative interpretation of the MAMI A1 line}

The MAMI A1 Collaboration reports that the observed line $p_{\pi^-} = 113.789 \pm 0.020_{\text{stat.}} \pm 0.112_{\text{syst.}}$~MeV/$c$ 
originates from $^{3}_{\Lambda}\mathrm{H} \to \pi^- + {}^{3}\mathrm{He}$ decay. This assignment
has recently been questioned by Gal~\cite{gal2026}, who proposed that the same sharp line may instead
originate from a different weak decay,

\begin{equation}
{}^{7}_{\Lambda}\mathrm{He}_{\text{g.s.}} \to \pi^- + {}^{7}\mathrm{Li}(E_x = 478~\text{keV}),
\label{eq:he7decay}
\end{equation}

This claim is motivated by the observation that $B_\Lambda({}^{3}_{\Lambda}\mathrm{H}) = 0.523$~MeV lies
more than $4\sigma$ from the pre-MAMI world-average value and is difficult to reconcile with
two decades of effective-field-theory expectations for this system. Working from the reported value MAMI A1 value $p_{\pi^-} = 113.789$~MeV/$c$, Ref. \cite{gal2026} computes a lambda binding energy of $B_\Lambda=5.84$ MeV. That paper further presents this as the binding energy of ${}^{7}_{\Lambda}\mathrm{He}$, a value that is completely different from the value of $B_\Lambda = 5.55 \pm 0.15$ ~MeV obtained from electroproduction measurements at Jefferson
Lab~\cite{gog2016}.

Using our methodology, we check whether the value of $B_\Lambda$ proposed in \cite{gal2026} does corresponds to the pion momentum observed by MAMI A1. The effective Q-value the negative pion decay of ${}^{7}_{\Lambda}\mathrm{He}_{\text{g.s.}}$ is 

\begin{equation}
Q_{\pi^-}^{\text{eff}} = Q_{\pi^-} - B_\Lambda({}^{7}_{\Lambda}\mathrm{He}) + S_p^{\text{eff}},
\label{eq:qeffhe7}
\end{equation}

where $Q_{\pi^-} = 37.841$~MeV is the same free-space Q-value used throughout this paper. The proton separation energy of the excited state of ${}^{7}\mathrm{Li}$ is obtained as follows:

\begin{equation}
S_p^{\text{eff}} = S_p({}^{7}\mathrm{Li},\text{g.s.}) - E_x = 9.973 - 0.478 = 9.495~\text{MeV}
\label{eq:speffhe7}
\end{equation}

where $S_p({}^{7}\mathrm{Li},\text{g.s.})=9.973$ MeV is the proton separation energy of the ground state of 7Li \cite{wan2021}, and $E_x=0.478$ MeV is the first excited state of 7Li that is considered in Ref. \cite{gal2026}. This relation follows directly from energy conservation. The equation to be solved by the Newton--Raphson method is

\begin{equation}
Q_{\pi^{-}}^{\mathrm{eff}} = \sqrt{\left(M(^{7}\mathrm{Li}^{*})c^{2}\right)^{2} + p_{\pi^{-}}^{2}c^{2}}
- M(^{7}\mathrm{Li}^{*})c^{2}
+ \sqrt{(m_{\pi^{-}}c^{2})^{2} + p_{\pi^{-}}^{2}c^{2}} - m_{\pi^{-}}c^{2}.
\label{eq:polynomial_7Li_general}
\end{equation}

Solving this equation for $p_{\pi^-}$, using $Q_{\pi^-}^{\text{eff}} = 41.496$ MeV, $M({}^{7}\mathrm{Li}^*)c^2 = 6534.311$~MeV and $m_{\pi^-}c^2 = 139.57039$~MeV, gives the results that are summarised in Table~\ref{tab:he7compare}. This table carries results for both $Q_{\pi^-}^{\text{eff}} = 41.496$ MeV (for the reassinged value $B_\Lambda = 5.84$ MeV) and $Q_{\pi^-}^{\text{eff}} = 41.786$ (for the JLab value $B_\Lambda = 5.55$ MeV).

\begin{table}[h]
\centering
\caption{Pion momentum for the $^{7}_{\Lambda}\mathrm{He} \to \pi^- + {}^{7}\mathrm{Li}^*$
channel, computed with the Newton--Raphson method for two values of
$B_\Lambda({}^{7}_{\Lambda}\mathrm{He})$.}
\label{tab:he7compare}
\begin{tabular}{lcc}
\hline
Quantity & Reassigned \cite{gal2026}) & JLab~\cite{gog2016} \\
\hline
$B_\Lambda({}^{7}_{\Lambda}\mathrm{He})$ (MeV) & 5.84 $\pm$ 0.07 & 5.55 $\pm$ 0.15 \\
$Q_{\pi^-}^{\text{eff}}$ (MeV) & 41.496 & 41.786 \\
$p_{\pi^-}$ (MeV/$c$) & 113.787 & 114.233 \\
$T_{\pi^-}$ (MeV) & 40.505 & 40.788 \\
$T_{{}^{7}\mathrm{Li}^*}$ (MeV) & 0.991 & 0.998 \\
\hline
\end{tabular}
\end{table}

It may be observed that the reassigned value proposed by Ref. \cite{gal2026} does reproduces the measured MAMI line
($p_{\pi^-} = 113.789 \pm 0.020_{\text{stat.}} \pm 0.112_{\text{syst.}}$~MeV/$c$). By contrast, using the accepted JLab value $B_\Lambda({}^{7}_{\Lambda}\mathrm{He}) = 5.55$~MeV predicts a line at $p_{\pi^-} = 114.233$~MeV/$c$, roughly $0.45$~MeV/$c$ above the momentum actually observed at MAMI. The exact kinematic formula reproduces identical values of negative pion momentum as the Newton--Raphson method. Reconciling the $^{7}_{\Lambda}\mathrm{He}$ interpretation with the MAMI line therefore requires a $B_\Lambda({}^{7}_{\Lambda}\mathrm{He})$ value higher than the standing JLab measurement: this is the origin of the internal tension in the alternative interpretation that Ref. \cite{gal2026} itself notes. 

Although the binding energy reassigned by Ref. \cite{gal2026} produces a pion momentum consistent with the observed MAMI line, this alternative interpretation faces significant experimental challenges. As demonstrated in the recent response by the A1 Collaboration~\cite{kin2026b}, the $^7_\Lambda$He hypothesis predicts a companion ground-state transition at $p_{\pi^-} \approx 114.5$ MeV/c with approximately twice the intensity of the observed peak. However, no statistically significant excess is observed in that region. Furthermore, Ref. \cite{gal2026}'s required $B_\Lambda(^7_\Lambda$He$) = 5.84 \pm 0.07$ MeV is in $\sim$2$\sigma$ tension with the direct spectroscopic measurement from the JLab HKS Collaboration ($B_\Lambda(^7_\Lambda$He$) = 5.55 \pm 0.10_{\rm stat} \pm 0.11_{\rm syst}$ MeV), for which no experimental justification has been provided. These quantitative inconsistencies strongly favour the original $^3_\Lambda$H assignment of the MAMI A1 peak.

\subsection{Neutral pion decay of hypertriton}

Neutral pion decay in $^3_\Lambda$H results in a $^3$H daughter nucleus and a $\pi^0$. Following a similar procedure, and using the constraint from momentum conservation $\vec{p}_{^3\text{H}} = -\vec{p}_{\pi^0}$, the equation for $p_{\pi^0}$ is given by

\begin{equation}
Q_{\pi^0}^{\text{eff}} = \sqrt{(M_{^3\text{H}}c^2)^2 + (p_{\pi^0}c)^2} - M_{^3\text{H}}c^2 + \sqrt{(m_{\pi^0}c^2)^2 + (p_{\pi^0}c)^2} - m_{\pi^0}c^2.
\end{equation}

Using the values $Q_{\pi^0}^{\text{eff}} = 46.875$ MeV,  $M(^3\text{H})c^2 = 2808.921$~MeV and $m_{\pi^0}c^2 = 134.9768$~MeV, this equation becomes

\begin{equation}
46.875 = \sqrt{2808.921^2 + p_{\pi^0}^2c^2} - 2808.921 + \sqrt{134.9768^2 + p_{\pi^0}^2c^2} - 134.9768,
\end{equation}

The Newton--Raphson numerical scheme is applied to solve for the pion momentum. The root $p_{\pi^0} = 118.129$~MeV/c was found. The convergence of the neutral pion momentum is shown in Table \ref{tab:pi0_convergence}. 

\begin{table}[h]
\centering
\caption{Newton--Raphson convergence for the neutral-pion decay momentum $p_{\pi^0}$.}
\label{tab:pi0_convergence}
\begin{tabular}{c c c}
\hline\hline
Iteration & $p$ (MeV/c)  \\
\hline
0 & 152.60248099  \\
1 & 120.20902153  \\
2 & 118.13966864  \\
3 & 118.12908256  \\
4 & 118.12908228  \\
5 & 118.12908228  \\
\hline\hline
\end{tabular}
\end{table}

From momentum conservation, the recoil momentum of $^3$H is $p_{^3\text{H}} = 118.129 $~MeV/c. The kinetic energy of each particle is given by
\begin{align}
T_{\pi^0} = \sqrt{(m_{\pi^0}c^2)^2 + (p_{\pi^0}c)^2} - m_{\pi^0}c^2 =44.392~\text{MeV} \\
T_{^3\text{H}} = \sqrt{(M_{^3\text{H}}c^2)^2 + (p_{\pi^0}c)^2} - M_{^3\text{H}}c^2 = 2.483~\text{MeV}
\end{align}

It may be observed that from the energy released, $Q_{\pi^0}^{\text{eff}} = 46.875$~MeV, the kinetic energy given to the neutral pion is 94.70\% of $Q_{\pi^0}^{\text{eff}}$), while the kinetic energy given to $^3$H (its recoil energy) is 5.30\% of $Q_{\pi^0}^{\text{eff}}$.

Just as in the negative-pion channel, the neutral-pion momentum can also be obtained from an exact relativistic two-body kinematic formula, without recourse to root-finding. Since $M^2-(m_1+m_2)^2$
and $M^2-(m_1-m_2)^2$ multiply out to $\left(M^2+m_2^2-m_1^2\right)^2-4M^2m_2^2$ for any daughter
masses $m_1,m_2$, the identical algebraic form used by Kamada et al.~\cite{kam1998} for the charged-pion
channel applies to the neutral-pion channel as well, with $M_{^3He}\to M_{^3H}$ and
$m_{\pi^-}\to m_{\pi^0}$:
\begin{align}
  k_{\pi^0} = \frac{\sqrt{\left(M_{\Lambda^3H}^2+M_{^3H}^2-m_{\pi^0}^2\right)^2-4M_{\Lambda^3H}^2M_{^3H}^2}}{2M_{\Lambda^3H}}.
  \label{eq:exact_pi0}
\end{align}
Using $M_{\Lambda^3H}=2990.773$ MeV (from $B_\Lambda=0.523$ MeV), $M_{^3H}=2808.921$ MeV, and
$m_{\pi^0}=134.9768$ MeV/$c^2$ in Eq.~\eqref{eq:exact_pi0} gives
\begin{align}
  k_{\pi^0} = \frac{\sqrt{\left(2990.773^2+2808.921^2-134.9768^2\right)^2-4(2990.773)^2(2808.921)^2}}{2(2990.773)} = 118.129 \text{ MeV}/c,
\end{align}
in exact agreement with the Newton--Raphson result of Table~\ref{tab:pi0_convergence}. Table~\ref{tab:pi0_comparison_formula}
extends this cross-check to the STAR and emulsion binding energies, confirming that the two methods agree
to within 0.001 MeV/$c$ in every case, and thereby validating the root-finding approach for the neutral-pion
channel exactly as was done for the charged-pion channel.

\begin{table}[htbp]
\centering
\caption{Comparison of the neutral-pion momentum from the Newton--Raphson root-finding method,
$p_{\pi^0}$, and the exact relativistic kinematic formula (Eq.~\eqref{eq:exact_pi0}), $p_{\pi^0}^{\rm formula}$,
for different $B_\Lambda$ measurements.}
\begin{tabular}{lccc}
\hline\hline
Quantity & MAMI A1 ($B_{\Lambda}=0.523$ MeV) & STAR ($B_{\Lambda}=0.406$ MeV) & Emulsion ($B_{\Lambda}=0.13$ MeV) \\
\hline
$M_{{}^3_\Lambda\mathrm{H}}$ (MeV)             & 2990.773 & 2990.890 & 2991.166 \\
$Q_{\pi^0}^{\rm eff}$ (MeV)                    & 46.875   & 46.992   & 47.268   \\
$p_{\pi^0}$ (MeV/$c$)                          & 118.129  & 118.296  & 118.689  \\
$p_{\pi^0}^{\rm formula}$ (MeV/$c$)            & 118.129  & 118.296  & 118.690  \\
$p_{^{3}\mathrm{H}}$ (MeV/$c$)                 & 118.129  & 118.296  & 118.689  \\
$T_{\pi^0}$ (MeV)                              & 44.392   & 44.502   & 44.762   \\
$T_{^{3}\mathrm{H}}$ (MeV)                     & 2.483    & 2.490    & 2.506    \\
\hline\hline
\end{tabular}
\label{tab:pi0_comparison_formula}
\end{table}

The kinematic trends in the neutral-pion decay mirror those of the negative-pion channel. The increase in $Q_{\mathrm{eff}}^{\pi^{0}}$ is again modest ($\sim 0.4$~MeV across the range of reported $B_{\Lambda}$ values). Furthermore, the relative variation in pion momentum remains small ($\sim 0.5\%$), showing that the predicted pion momentum is only weakly sensitive to the spread in current $B_{\Lambda}$ measurements. This again highlights the necessity for high precision in experimental measurements.

\section{Conclusion}

Neutral-pion momentum in the two-body mesonic decay of the hypertriton, $^3_\Lambda\mathrm{H} \to {}^3\mathrm{H} + \pi^0$, has been determined for the first time through a robust root-finding approach based on four-momentum conservation. Employing the Newton--Raphson algorithm and the high-precision $\Lambda$ binding energy $B_\Lambda = 0.523 \pm 0.013\,(\text{stat.}) \pm 0.075\,(\text{syst.})$\,MeV from the MAMI A1 Collaboration, we obtain a predicted neutral-pion momentum of $p_{\pi^0} = 118.129$\,MeV/$c$. Furthermore, we showed that the root-finding method agrees with the exact relativistic two-body kinematic relation  in both channels.

The method was thoroughly validated on the well-measured charged-pion channel $^3_\Lambda\mathrm{H} \to {}^3\mathrm{He} + \pi^-$, where it reproduces the experimental momentum $p_{\pi^-} = 113.789$\,MeV/$c$ with a remarkable relative deviation of only $0.0009\%$. This agreement underscores the reliability and numerical efficiency of the technique. In both channels, the pion carries approximately 94.6--94.7\% of the available effective $Q$-value, while the recoiling trinucleon absorbs the remainder ($\sim$5.3--5.4\%).

Comparative analyses using alternative $B_\Lambda$ values from the STAR Collaboration and the older emulsion experiment reveal only modest variations ($\sim$0.5\%) in the predicted pion momenta, highlighting the relative insensitivity of these kinematic observables to current uncertainties in the hypertriton binding energy. Such robustness reinforces the value of decay-pion spectroscopy as a precise probe of hypernuclear structure.

This work fills an important gap by providing a concrete prediction for the neutral-pion momentum, which remains experimentally inaccessible directly with present techniques due to the detection challenges associated with $\pi^0 \to \gamma\gamma$ decays. The study demonstrates the power of root-finding techniques for extracting hypernuclear kinematics where direct measurement remains difficult and also for three-body decay channels where the exact kinematic formula breaks down. 

\bibliographystyle{unsrtnat}
\bibliography{mesonic_hypertriton}
\end{document}